\documentclass[11pt,showkeys,aps,pra]{revtex4}
\usepackage[utf8]{inputenc}
\usepackage[draft]{graphics}
\usepackage{graphicx}          	 
\usepackage{bm}                 
\usepackage{amsmath}             
\usepackage{amssymb}
\usepackage{amsfonts}             
\usepackage{verbatim}           
\usepackage{amsthm}             
\usepackage{mathtools}

\begin{document}

\title{One-parameter Fisher-Rényi complexity: Notion and hydrogenic applications}

\author{I.V. Toranzo$^{a,b}$, P. Sánchez-Moreno$^{a,c}$, {\L}ukasz Rudnicki$^{d,e}$ and J.S. Dehesa$^{a,b}$}
\affiliation{
$^a$Instituto {\em Carlos I} de F\'isica Te\'orica y Computacional, Universidad de Granada, 18071-Granada, Spain\\
$^b$Departamento de F\'isica At\'omica, Molecular y Nuclear, Universidad de Granada, 18071-Granada, Spain\\
$^c$Departamento de Matem\'atica Aplicada, Universidad de Granada, 18071-Granada, Spain\\
$^d$Institute for Theoretical Physics, University of Cologne, Z\"ulpicher Stra{\ss}e 77,
D-50937, Cologne, Germany\\
$^e$Center for Theoretical Physics, Polish Academy of Sciences, Aleja Lotnik{\'o}w 32/46,
PL-02-668 Warsaw, Poland}
%
%

\begin{abstract}
In this work the one-parameter Fisher-Rényi measure of complexity for general $d$-dimensional probability distributions is introduced and its main analytic properties are discussed. Then, this quantity is determined for the hydrogenic systems in terms of the quantum numbers of the quantum states and the nuclear charge.
\end{abstract}

\keywords{Information theory; Fisher information; Shannon entropy; Rényi entropy; Fisher-Rényi complexity; hydrogenic systems}

\maketitle

\section{Introduction}
We all have an intuitive sense of what \textit{complexity} means. In the last two decades an increasing number of efforts have been published \cite{lopezruiz,gellmann1995,gellmann1996,badii,gregersen,gellmann,frieden_04, sen2012,zuchowski,seitz,bawden,lukasz} to refine our intuitions about complexity into precise, scientific concepts, pointing out a large amount of open problems. Nevertheless there is not a consensus for the term \textit{complexity} nor whether there is a simple core to \textit{complexity}. Contrary to the Boltzmann-Shannon entropy which is ever increasing according to the second law of thermodynamics, the complexity seems to behave very differently.   Various precise, widely applicable, numerical and analytical proposals (see e.g., \cite{shiner,kolmogorov,VignFS,catalan_pre02,lloyd,yamano_pa04,  plastino,pipek,lopez,lopezr,angulo_pla08,romera_1,romera08,romera09,dehesa_1,antolin_ijqc09,psanchez,vedral} and the monograph \cite{sen2012}) have been done but they are yet very far to appropriately formalize the intuitive notion of complexity \cite{psanchez,lukasz}. The latter suggests that complexity should be minimal at either end of the scale. However, a complexity quantifier to take into account the completely ordered and completely disordered limits (i.e., perfect order and maximal randomness, respectively) and to describe/explain the maximum between them is not known up until now. \\

Recently, keeping in mind the fundamental principles of the density functional theory, some statistical measures of complexity have been proposed to quantify the degree of structure or pattern of finite many-particle systems in terms of their single-particle density, such as the Crámer-Rao  \cite{dehesa_1,antolin_ijqc09}, Fisher-Shannon \cite{VignFS,angulo_pla08,romera_1} and LMC (López-ruiz, Mancini and Calvet) \cite{lopezruiz,catalan_pre02} complexities and some modifications of them  \cite{pipek,lopez,lopezr,romera08,romera09,psanchez}. They are composed by a two-factor product of entropic measures of Shannon \cite{shannon_49}, Fisher \cite{fisher,frieden_04}  and Rényi \cite{renyi_70} types. Most interesting for quantum systems are those which involve the Fisher information (namely, the Crámer-Rao and the Fisher-Shannon complexities, and their modifications \cite{romera08,antolin09,romera09}), mainly because this is by far the best entropy-like quantity to  take into account the inherent fluctuations of the quantum wave functions by quantifying the gradient content of the single-particle density of the systems. \\

The objetive of this article is to extend and generalize these Fisher-information-based measures of complexity by introducing a new complexity quantifier, the one-parameter Fisher-Rényi complexity, to discuss its properties and to apply it to the main prototype of Coulombian systems, the hydrogenic system. This notion is composed by two factors: a $\lambda$-dependent Fisher information (which quantifies various aspects of the quantum fluctuations of the physical wave functions beyond the density gradient, since it reduces to the standard Fisher information for $\lambda=1$) and the Rényi entropy of order $\lambda$ (which measures various facets of the spreading or spatial extension of the density beyond the celebrated Shannon entropy which corresponds to the limiting case $\lambda \to 1$). \\

The article is structured as follows. In Section I we introduce the notion of one-parameter Fisher-Rényi measure of complexity. In Section II we discuss the main analytical properties of this complexity, showing that it is bounded from below, invariant under scaling transformations and monotone. In addition the near-continuity and the invariance under replications are also discussed.  In Section III, we apply the new complexity measure to the hydrogenic systems. Finally some concluding remarks are given.

\section{One-parameter Fisher-Rényi complexity measure}

In this section the notion of one-parameter Fisher-Rényi complexity $C_{FR}^{(\lambda)}[\rho]$ of a $d$-dimensional probability density is introduced and its main analytic properties are discussed.  This quantity is composed by two entropy-like factors of local (the one-parameter Fisher information of Johnson and Vignat \cite{vignat}, $\tilde{F}_\lambda[\rho]$) and global (the $\lambda$-order Rényi entropy power \cite{toscani}, $N_{\lambda}[\rho]$) characters.

\subsection{The notion}

The one-parameter Fisher-R\'enyi complexity measure $C_{FR}^{(\lambda)}[\rho]$ of the probability density $\rho(x),\, x = (x_1, x_2,\ldots,x_d) \in \mathbb{R}^{d}$, is defined by
\begin{equation}
\label{eq:def}
C_{FR}^{(\lambda)}[\rho] = D_\lambda^{-1} \tilde{F}_\lambda[\rho]N_\lambda[\rho], \quad \lambda >\max\left\{\frac{d-1}{d},\frac{d}{d+2} \right\},
\end{equation}
where $D_\lambda$ is the normalization factor given as 
\begin{equation}
\label{eq:normfactor}
D_\lambda = \left\{
\begin{array}{ll}
2\pi d\frac{\lambda^{-1}}{\lambda-1}\left(\frac{\Gamma\left(\frac{\lambda}{\lambda-1}\right)}{\Gamma\left(\frac{d}{2}+\frac{\lambda}{\lambda-1}\right)}\right)^{\frac{2}{d}}\left(\frac{(d+2)\lambda-d}{2\lambda}\right)^{\frac{2+d(\lambda-1)}{d(\lambda-1)}}, & \lambda>1\\
 2\pi d\frac{\lambda^{-1}}{1-\lambda}\left(\frac{\Gamma\left(\frac{1}{1-\lambda}-\frac{d}{2}\right)}{\Gamma\left(\frac{1}{1-\lambda}\right)}\right)^{\frac{2}{d}}\left(\frac{(d+2)\lambda-d}{2\lambda}\right)^{\frac{2+d(\lambda-1)}{d(\lambda-1)}}, & \max\left\{\frac{d-1}{d},\frac{d}{d+2}\right\}<\lambda<1.
\end{array}
\right.
\end{equation}
This purely numerical factor is necessary to let the minimal value of the complexity be equal to unity, as explained below in paragraph 2.2.1. The $\tilde{F}_\lambda[\rho]$ denotes the (scarcely known) $\lambda$-weighted Fisher information \cite{vignat} defined by
\begin{equation}
\label{eq:gen_fish1}
\tilde{F}_\lambda[\rho]=\left(\int_{\mathbb{R}^{d}} \rho^\lambda(x)\, dx \right)^{-1} \int_{\mathbb{R}^{d}} |\rho^{\lambda-2}(x) \nabla\rho(x)|^2\rho(x)\, dx,
\end{equation}
(which, for $\lambda= 1$, reduces to the standard Fisher information $F[\rho]=\int_{\mathbb{R}^{d}} \frac{|\nabla \rho|^{2}}{\rho}\,dx$), being $dx$ the $d$-dimensional volume element. Finally, the symbol
$N_{\lambda}[\rho]$ denotes the $\lambda$-Rényi entropy power (see e.g., \cite{toscani}) given as
\begin{equation}
    N_{\lambda}[\rho] = \left\{\begin{array}{ll}
    \left(\int_{\mathbb{R}^{d}}  \rho^\lambda(x) \,dx \right)^{\frac{\mu}{d}\frac{1}{1-\lambda}}& \text{if} \quad \lambda \neq 1, 0<\lambda <\infty, \\[2mm]
    e^{\frac{2}{d}S[\rho]}& \text{if} \quad \lambda = 1,\\
    \end{array} \right.
  \end{equation}
  where $\mu=2+d(\lambda-1)$ and $S[\rho] := - \int_{\mathbb{R}^{d}}  \rho(x)\ln \rho(x)\,dx$ is the  Shannon entropy \cite{shannon_49}.\\

The complexity measure $C_{FR}^{(\lambda)}[\rho]$ has a number of conceptual advantages with respect to the Fisher-information-based measures of complexity previously defined; namely, the Crámer-Rao and Fisher-Shannon  complexity and their modifications. Indeed, it quantifies the combined balance of different ($\lambda$-dependent) aspects  of both the fluctuations and the spreading or spatial extension of the single-particle density $\rho$ in such a way that there is no dependence on any specific point of the system's region. The Crámer-Rao complexity \cite{dehesa_1,antolin_ijqc09} (which is the product of the standard Fisher information $F[\rho]$ mentioned above and the variance $V[\rho] = \langle r^2\rangle - \langle r\rangle^2$) measures a single aspect of the fluctuations (namely, the density gradient) together with the concentration of the probability density around the centroid $\langle r \rangle$. The Fisher-Shannon complexity \cite{VignFS,angulo_pla08,romera_1}, defined by $C_{FS}[\rho]= F[\rho]\times e^{\frac{2}{d}S[\rho]}$, quantifies the density gradient jointly with a single aspect of the spreading given by the Shannon entropy $S[\rho]$ mentioned above. A modification of the previous measure by use of the Rényi entropy $R_{\lambda}[\rho] = \frac{1}{1-d} \ln \int_{\mathbb{R}^{d}}  \rho^\lambda(x) \,dx$ instead of the Shannon entropy, the Fisher-Rényi product of complexity-type, has been recently introduced \cite{romera08,antolin09,romera09}; it measures the gradient together with various aspects of the spreading of the density.

\subsection{The properties}
Let us now discuss some properties of this notion: bounding from below, invariance under scaling transformations, monotonicity, behavior under replications and near continuity.
\begin{enumerate}
\item \textbf{Lower bound.} The Fisher-R\'enyi complexity measure $C_{FR}^{(\lambda)}[\rho]$ fulfills the inequality
\begin{equation}
C_{FR}^{(\lambda)}[\rho]\ge 1
\label{eq:toscani_inequality}
\end{equation}
(for $\lambda>\max\left\{\frac{d-1}{d},\frac{d}{d+2}\right\}$, with $\lambda\neq 1$), and the minimal complexity occurs, as implicitly proved by Savaré and Toscani \cite{toscani}, if and only if the density has the following generalized Gaussian form 
\begin{equation}
\label{eq:density_Bp}
\mathcal{B}_{\lambda}(x) = \left\{\begin{array}{ll}
(C_{\lambda}-|x|^{2})_{+}^{\frac{1}{\lambda-1}} , & \lambda>1\\
(C_{\lambda}+|x|^{2})^{\frac{1}{\lambda-1}} , & \lambda<1
\end{array}\right.
\end{equation}
where  $(x)_{+}=\max\{x,0\}$ and $C_{\lambda}$ is the normalization constant given by
\begin{equation}
\label{eq:norm}
C_{\lambda} = A_{\lambda}^{-\frac{2(\lambda-1)}{d(\lambda-1)+2}},
\end{equation}
with 
\begin{equation*}
A_\lambda = \left\{
\begin{array}{ll}
\pi^{d/2}\frac{\Gamma\left(\frac{\lambda}{\lambda-1}\right)}{\Gamma\left(\frac{d}{2}+\frac{\lambda}{\lambda-1}\right)}, & \lambda>1 \\
\pi^{d/2}\frac{\Gamma\left(\frac{1}{1-\lambda}-\frac{d}{2}\right)}{\Gamma\left(\frac{1}{1-\lambda}\right)}, & \frac{d}{d+2}<\lambda<1\\
\end{array}
\right.
\end{equation*}
Thus, the complexity measure $C_{FR}^{(\lambda)}(\rho)$ has a universal lower bound of minimal complexity, that is achieved for the family of densities $\mathcal{B}_{\lambda}(x)$. 

\item \textbf{Invariance under scaling and translation transformations}. The complexity measure $C_{FR}^{(\lambda)}(\rho)$ are scaling and translation invariant in the sense that
\begin{equation}
\label{eq:invsca}
C_{FR}^{(\lambda)}[\rho_{a,b}] = C_{FR}^{(\lambda)}[\rho], \quad \forall \lambda,
\end{equation}
where $\rho_{a,b}(x) = a^{d}\rho(a(x-b))$, with $a\in \mathbb{R}$ and $b\in\mathbb{R}^{d}$. To prove this property we follow the lines of Savaré and Toscani \cite{toscani}. First we calculate the generalized Fisher information of the transformed density, obtaining
\begin{eqnarray*}
\tilde{F}_{\lambda}[\rho_{a,b}] &=& \left(\int_{\mathbb{R}^{d}}a^{d\lambda}\rho^{\lambda}(a(x-b))\,dx \right)^{-1}\\
& \times& \int_{\mathbb{R}^{d}} a^{2d(\lambda-2)}\rho^{2(\lambda-2)}(a(x-b))|a^{d+1}[\nabla\rho](a(x-b))|^{2}a^{d}\rho(a(x-b))\,dx \\
&=& a^{d (\lambda -1)+2}\left(\int_{\mathbb{R}^{d}}\rho^{\lambda}(y)\,dy \right)^{-1}\int_{\mathbb{R}^{d}} \rho^{2\lambda-4}(y)|\nabla\rho(y)|^{2}\rho(y)\,dy \\
&\equiv & a^{d (\lambda -1)+2}\tilde{F}_{\lambda}[\rho], \quad \forall \lambda
\end{eqnarray*}
Note that in writing the first equality we have used that $$|\nabla \rho_{a,b}(x)|^{2} = |a^{d+1}[\nabla \rho](a(x-b))|^{2}.$$ Then, we determine the value of the $\lambda$-entropy power of the density $\rho_{a,b}(x)$ which turns out to be equal to
\begin{eqnarray*}
N_{\lambda}[\rho_{a,b}] &=& \left(\int_{\mathbb{R}^{d}} a^{d\lambda} \rho^{\lambda}(a(x-b))\,dx \right)^{\frac{2+d(\lambda-1)}{d(1-\lambda)}}\\
&=& \left( a^{d(\lambda-1)} \int_{\mathbb{R}^{d}}  \rho^{\lambda}(y)\,dy \right)^{\frac{2+d(\lambda-1)}{d(1-\lambda)}}\\
&\equiv& a^{-d(\lambda-1)-2}N_{\lambda}[\rho], \quad \forall \lambda
\end{eqnarray*}
In particular, we have
\begin{eqnarray*}
N_{1}[\rho_{a,b}] &=& \exp\left[-\frac{2}{d}\int_{\mathbb{R}^{d}} a^{d} \rho^{\lambda}(a(x-b))\ln[a^{d}\rho^{\lambda}(a(x-b))]\,dx \right]\\
&=& \exp\left[-\frac{2}{d}\int_{\mathbb{R}^{d}} \rho(y)\ln[a^{d}\rho(y)]\,dy \right]\\
&=& \exp\left[-\frac{2}{d} \left( d\ln a + S[\rho] \right)  \right] \\
&\equiv& a^{-2} N_{1}[\rho],
\end{eqnarray*}
Finally, from Eq. (\ref{eq:def}) and the values of $\tilde{F}_{\lambda}[\rho_{a,b}]$ and $N_{\lambda}[\rho_{a,b}]$ just found, we readily obtain the wanted invariance (\ref{eq:invsca}).

\item \textbf{Monotonicity}.
%
%
The existence of a non-trivial operation with interesting properties under which a complexity measure is monotonic \cite{lukasz} is a valuable property of the measure in question from the axiomatic point of view. To show the monotonic behavior of the Fisher-Rényi complexity $C_{FR}^{(\lambda)}(\rho)$ we make use of the so-called \textit{rearrangements}, which represent a useful tool in the theory of functional analysis and, among other applications, have been used to prove relevant inequalities such as Young’s inequality with sharp constant. \\
Two of the main properties of rearrangements is that they preserve the $L^{p}$ norms, which implies that the rearrangements of a probability density give rise to another probability density, and that they make everything spherically symmetric. The second feature makes the rearrangement operation relevant for quantification of statistical complexity \cite{lukasz}, since a spherically symmetric variant of a probability density can in an atomic context be viewed as less complex. Then, we introduce the definition of this operation as well as its effects over the entropic quantities that make up our complexity measure. \\
Let $f$ be a real-valued function, $f:\mathbb{R}^{n} \rightarrow [0,\infty)$ and $A_{t} =\{x:f(x)\geq t \}$. The symmetric decreasing rearrangement of $f$ is defined as
\begin{equation}
\label{eq:symmrear}
f^{*}(x) = \int_{0}^{\infty} \chi_{\{x\in A^{*}_{t} \}}\, dt\, ,
\end{equation}
with $\chi_{\{x\in A^{*}_{t} \}}=1$ if $x\in A^{*}_{t} $ and $0$ otherwise. $A_{t}$ represents the super-level set of the function $f$ and $A^{*}$ (which denotes the symmetric rearrangement of a set $A\subset\mathbb{R}^{n}$) is the Euclidean ball centered at $0$ such as $Vol(A^{*}) = Vol(A)$.\\
The central idea of this transformation is to build up $f^{*}$ from the rearranged super-level sets in the same manner that $f$ is built from its super-level sets. As a by-product from its construction, $f^{*}$ turns out to be a \textit{spherically symmetric decreasing} function (i.e. $f^{*}(x)=f^{*}(|x|)$ and moreover $f^{*}(b)<f^{*}(a)\,\, \forall b>a$, where $a,b\in A_{t}^{*}$) which means that for any function $f:\mathbb{R}^{n}\rightarrow [0,\infty)$ and all $t\geq 0$
\begin{equation}
\label{eq:fulfil}
\{x:f(x)>t\}^{*} = \{x:f^{*}(x)>t\},
\end{equation}
or in other words, that for any measurable subset $B\subset [0,\infty)$, the volume of the sets $\{x:f(x)\in B \}$ and $\{x:f^{*}(x)\in B \}$ are the same.

It is known \cite{madiman} that under this transformation and for any $p\in [0,1)\cup(1,\infty]$ the Rényi and Shannon entropies remain unchanged, i.e.
\begin{equation}
\label{eq:rearRS}
R_{p}[\rho] = R_{p}[\rho^{*}] , \quad S[\rho] = S[\rho^{*}]
\end{equation}
if both $S[\rho]$ and $S[\rho^{*}]$ are well defined, where $\lim_{p\to 1} R_{p}[\rho] = S[\rho]$.
The invariance of the Rényi entropy follows from the preservation of the $L^{p}$ norms via rearrangements and the proof of the invariance of the Shannon entropy is done in \cite{madiman}. Moreover, Wang and Madiman \cite{madiman} consider the Fisher information, finding that the standard Fisher information decreases monotonically under rearrangements, i.e.
\begin{equation}
\label{eq:rearFish}
F[\rho] \geq  F[\rho^{*}]. 
\end{equation}

Let us now consider the biparametric Fisher-like information, $I_{\beta,q}[f]$, of a probability density function $f(x)$ which is defined \cite{bercher} by
\begin{equation}
	I_{\beta,q}[f] = \int_{\mathbb{R}^{d}} f^{\beta(q-1)+1}(x) \Big(\frac{|\nabla f(x)|}{f(x)}\Big)^{\beta}f(x)\, dx
\end{equation}
with $q \geq 0,\, \beta > 1$. Then one notes that the one-parameter Fisher information, $\tilde{F}_{\lambda}[\rho]$, given by (\ref{eq:gen_fish1}) can be expressed in terms of the previous quantity with $\beta=2$ and $q\equiv \lambda$ as
\begin{equation}
\label{eq:genFish}
\tilde{F}_{\lambda}[\rho] = \frac{\int_{\mathbb{R}^{d}} |\rho^{\lambda-2}(x) \nabla\rho(x)|^2\rho(x)\, dx}{\int_{\mathbb{R}^{d}} \rho^\lambda(x)\, dx}=\frac{I_{2,\lambda}[\rho]}{N_{\lambda}[\rho]^{\frac{\mu}{d}(1-\lambda)}}.
\end{equation}
On the other hand, considering the transformation $\rho = u(x)^{k}$ with $k = \frac{\beta}{\beta(q-1)+1}$, the biparametric Fisher information becomes
\begin{equation}
\label{eq:bipFish}
I_{\beta,q} = \int_{\mathbb{R}^{d}}|\nabla u(x)|^{\beta}\, dx
\end{equation}
also known as the $\beta$-Dirichlet energy of $u(x)$. If $k = 2$, note that the function $u(x)$ corresponds to a quantum-mechanical wave function. By using the  symmetric decreasing rearrangement to the density function $\rho$, the well-known  Pólya-Szeg\"o inequality states that 
\begin{equation}
\label{eq:PSineq}
I_{\beta,q}[\rho]=\int_{\mathbb{R}^{d}}|\nabla u|^{\beta} \geq I_{\beta,q}[\rho^{*}]= \int_{\mathbb{R}^{d}}|\nabla u^{*}|^{\beta} ,
\end{equation}
which implies that the minimizer of the left side is necessarily radially symmetric and decreasing, so the extremal function belongs to the subset of radially symmetric probability densities, and is represented by the generalized Gaussian given in (\ref{eq:density_Bp}). Now by taking into account (\ref{eq:genFish}) and the invariance of the Rényi entropy (and therefore  the Rényi entropy power, $N_{\lambda}[\rho]$) upon rearrangements one obtains the monotonic behavior of $\tilde{F}_{\lambda}[\rho]$ as
\begin{equation}
\label{eq:Ftmon}
\tilde{F}_{\lambda}[\rho] = \frac{I_{2,\lambda}[\rho]}{N_{\lambda}[\rho]^{\frac{\mu}{d}(1-\lambda)}} \geq \tilde{F}_{\lambda}[\rho^{*}] = \frac{I_{2,\lambda}[\rho^{*}]}{N_{\lambda}[\rho^{*}]^{\frac{\mu}{d}(1-\lambda)}},  \end{equation}
Finally, this observation together with (\ref{eq:def}) allows us to obtain the monotonic behavior of this complexity measure $C_{FR}^{(\lambda)}(\rho)$ proved by rearrangements, i.e.
\begin{equation}
\label{eq:monComp}
C_{FR}^{(\lambda)}(\rho) \geq C_{FR}^{(\lambda)}(\rho^{*}),
\end{equation}
where the inequality is saturated for the generalized Gaussian, $\rho(x)=\mathcal{B}_{\lambda}(x)$, which also means that the symmetric rearrangement of a generalized Gaussian gives another generalized Gaussian, i.e. rearrangements preserve this subset of radially symmetric probability densities $\mathcal{B}^{*}_{\lambda}(x) = \mathcal{B}_{\lambda'}(x)$.
\vspace{3mm}

\item \textbf{Behavior under replications}. Let us now study the behavior of the Fisher-Rényi complexity $C_{FR}^{(\lambda)}(\rho)$ under $n$ replications. We have found that for one-dimensional densities $\rho(x),\, x\in\mathbb{R}$ with bounded support, this complexity measure behaves as follows:
\begin{equation}
\label{eq:repli}
C_{FR}[\tilde{\rho}] = n^2 C_{FR}[\rho],	
\end{equation}
where the density $\tilde{\rho}$ representing $n$ replications of $\rho$ is given by
\[
\tilde{\rho}(x) = \sum_{m=1}^n\rho_m(x);\quad \rho_m(x)= n^{-\frac12} \rho\left(n^\frac12(x-b_m)\right),
\]
where the points $b_m$ are chosen such that the supports $\Lambda_m$ of each density $\rho_m$ are disjoints. Then, the integrals
\begin{multline*}
\int_\Lambda |(\tilde{\rho}(x))^{\lambda-2}\tilde{\rho}'(x)|^2\tilde{\rho}(x)dx
=\sum_{m=1}^n\int_{\Lambda_m} |(\rho_m(x))^{\lambda-2}\rho_m'(x)|^2 \rho_m(x)dx\\
=\sum_{m=1}^n n^{-\lambda+1} \int_\Lambda |(\rho(y))^{\lambda-2}\rho'(y)|^2 \rho(y)dy
=n^{-\lambda+2} \int_\Lambda |(\rho(y))^{\lambda-2}\rho'(y)|^2 \rho(y)dy,
\end{multline*}
and
\begin{multline*}
\int_\Lambda (\tilde{\rho}(x))^\lambda dx
= \sum_{m=1}^n \int_{\Lambda_m} (\tilde{\rho}_m(x))^\lambda dx
= \sum_{m=1}^n n^{-\frac{\lambda+1}{2}} \int_\Lambda (\rho(y))^\lambda dy
= n^{-\frac{\lambda-1}{2}} \int_\Lambda (\rho(y))^\lambda dy,
\end{multline*}
where the change of variable $y= n^\frac12 (x-b_m)$ has been performed.

Thus, the two entropy factors (the generalized Fisher information and the Rényi entropy power) of the Fisher-Rényi measure $C_{FR}^{(\lambda)}(\rho)$ gets modified as
\begin{equation}
\tilde{F}_\lambda[\tilde{\rho}] = n^{\frac{3-\lambda}{2}} \tilde{F}_\lambda[\rho], \quad \quad N_\lambda[\tilde{\rho}] = n^{\frac{\lambda+1}{2}} N_\lambda[\rho], 	
\end{equation}
so that from these two values and (\ref{eq:def}) we finally have the wanted behavior (\ref{eq:repli}) of the Fisher-Rényi complexity under $n$ replications. Although this has been proved in the one dimensional case, similar arguments hold for general dimensional densities.

\item \textbf{Near-continuity behavior}. Let us now illustrate that the Fisher-Rényi complexity is not near continuous by means of a one-dimensional counter-example. Recall first that a functional $G$ is near continuous if for any $\epsilon>0$ exist $\delta>0$ such that, if two densities $\rho$ and $\tilde{\rho}$ are $\delta$-neighboring  (i.e., the Lebesgue measure of the points satisfying $|\rho(x)-\tilde{\rho}(x)|\ge \delta$ is zero), then $|G[\rho]-G[\tilde{\rho}]|<\epsilon$. Now, let us consider the $\delta$-neighboring densities
\[
\rho(x)=\frac{2}{\pi}\left\{
\begin{array}{ll}
\sin^2(x),& -\pi\le x\le 0,\\
0, & \text{elsewhere},
\end{array}
\right.
\]
and
\[
\tilde{\rho}(x)=\frac{2}{\pi(1+\delta^6)}\left\{
\begin{array}{ll}
\sin^2(x),& -\pi\le x\le 0,\\
\delta \sin^2\left(\frac{x}{\delta^5}\right),& 0<x\le \delta^5\pi,\\
0, & \text{elsewhere}.
\end{array}
\right.
\]

Due to the increasing oscillatory behaviour of $\tilde{\rho}$ for $x\in(0,\delta^5\pi)$ as $\delta$ tends to zero, the generalized Fisher information $\tilde{F}$ grows rapidly as $\delta$ decreases, while the R\'enyi entropy power tends to a constant value. Then, the more similar $\rho$ and $\tilde{\rho}$ are, the more different are their values of $C^{(\lambda)}_{FR}$. Therefore, the Fisher-Rényi complexity measure is not near continuous.

\end{enumerate}

\section{The hydrogenic application}
 
In this section we determine the one-parameter Fisher-Rényi complexity measure $C^{(\lambda)}_{FR}$, given by (\ref{eq:def}), for the probability density of hydrogenic atoms consisting of an electron bound by the Coulomb potential, $V(r) = - \frac{Z}{r}$, where $Z$ denotes the nuclear charge, $r \equiv |\vec{r}| = \sqrt{\sum_{i=1}^3 x_i^2}$ and the position vector $\vec{r}  =  (x_1 , x_2, x_3)$ is given in spherical polar coordinates as $(r,\theta,\phi)      \equiv
(r,\Omega)$, $\Omega\in S^2$. Atomic units are used. The hydrogenic states are well known to be characterized by the three quantum numbers \{$n, l, m$\}, with $n=0,1,2,\ldots$, $l=0,1,\ldots, n-1$ and $m=-l,-l+1,\ldots,l$. They have the
  energies $E_{n}=-\frac{Z^{2}}{2n^{2}}$, and the corresponding quantum probability densities are given by
 \begin{equation}
 \label{eq:qpd}
\rho_{n,l,m}(\vec{r}) = \rho_{n,l}(\tilde{r})\,\,\Theta_{l,m}(\theta,\phi)
 \end{equation}
 where $\tilde{r}=\frac{2 Z}{n}r $, and the symbols $\rho_{n,l}(\tilde{r})$ and $\Theta_{l,m}(\theta,\phi)$ are the radial and angular parts of the density, which are given by 
  \begin{equation}
  \label{eq:ronl}
  \rho_{n,l}(\tilde{r}) = \frac{4Z^{3}}{n^{4}}\frac{\omega_{2l+1}(\tilde{r})}{\tilde{r}}[\widehat{L}_{n-l-1}^{(2l+1)}(\tilde{r})]^{2}	
  \end{equation}
and 
  \begin{equation}
  \Theta_{l,m}(\theta,\phi) = |Y_{l,m}(\theta,\phi)|^{2},	
  \end{equation}
  respectively. In addition, $\widehat{L}_{n}^{\alpha}(x)$ denotes the orthonormal Laguerre polynomials \cite{nist} with respect to the weight function $\omega_{\alpha}=x^{\alpha}e^{-x}$ on the interval $[0,\infty)$, and $Y_{l,m}(\theta,\phi)$  are the well-known spherical harmonics which can be expressed in terms of the Gegenbauer polynomials, $C_{n}^{m}(x)$ via
 \begin{equation}
 \label{eq:harm_legend}
Y_{l,m}(\theta,\phi) = \left(\frac{(l+\frac{1}{2})(l-|m|)![\Gamma(|m|+\frac{1}{2})]^{2}}{2^{1-2|m|}\pi^{2}(l+|m|)!} \right)^{\frac{1}{2}}e^{im\phi}(\sin\theta)^{|m|}C_{l-|m|}^{|m|+\frac{1}{2}}(\cos\theta),
  \end{equation}
where $0\leq \theta \leq \pi$ and $0\leq \phi \leq 2\pi$.  Let us now compute the complexity measure $C^{(\lambda)}_{FR}[\rho_{n,l,m}]$ of the hydrogenic probability density which, according to (\ref{eq:def}), is given by 
\begin{eqnarray}
\label{eq:exp1}
  C^{(\lambda)}_{FR}[\rho_{n,l,m}] &=&  D_\lambda^{-1} \tilde{F}_\lambda[\rho_{n,l,m}]N_\lambda[\rho_{n,l,m}]
  \equiv D_\lambda^{-1}I_{1}I_{2}^{2 \left(\frac{1}{3 (1-\lambda )}-1\right)}, 
\end{eqnarray} 
where $D_\lambda$ is the normalization constant given by (\ref{eq:normfactor}) and the symbols $I_{1}$ and $I_{2}$ denote the integrals
 \begin{eqnarray}
 \label{eq:1A}
 I_{1} &=& \int |\left[\rho_{n,l,m}(\vec{r})\right]^{\lambda-2}\,\,\nabla\rho_{n,l,m}(\vec{r})|^{2}\,\rho_{n,l,m}(\vec{r})\, d^3\vec{r}=\int \left[\rho_{n,l,m}(\vec{r})\right]^{2\lambda-3}\,\,|\nabla\rho_{n,l,m}(\vec{r})|^{2}\,d^3\vec{r}, \\
 I_{2} &=&  \int \left[\rho_{n,l,m}(\vec{r})\right]^{\lambda}\, d^3\vec{r} = \int_{0}^{\infty} \left[\rho_{n,l}(\tilde{r})\right]^{\lambda}\,r^{2}\,dr \int_{\Omega}\left[\Theta_{l,m}(\theta,\phi)\right]^{\lambda}\,d\Omega, 
\end{eqnarray} 
which can be solved by following the lines indicated in Appendix \ref{integrals:app}.

In the following, for simplicity and illustration purposes, we focus our attention on the computation of the complexity measure for two large, relevant classes of hydrogenic states: the $(ns)$ and the circular $(l=m=n-1)$ states. 
\begin{enumerate}
\item Generalized Fisher-Rényi complexity of hydrogenic $(ns)$ states.

In this case, $\Theta_{0,0}(\theta,\phi) = |Y_{0,0}(\theta,\phi) |^{2}=\frac{1}{4\pi}$ so that the three angular integrals can be trivially determined, and the radial integrals simplify as
\begin{eqnarray}
\label{eq:redu_I1rad}
I_{1a}^{(rad)} &=& \frac{2^{4 \lambda -3} Z^{6 \lambda -4}}{ n^{10 \lambda-6 }}(2 \lambda -1)^{-1}\mathcal{G}(n,0,\lambda)\\
I_{1b}^{(rad)}\,& =& \frac{2^{4\lambda-3}Z^{6\lambda-4}}{n^{10\lambda-6}}(2\lambda-1)^{-1}\Phi_{0}\left(0,0,2(2\lambda-1),\{n-1\}, \{1\}; \left\{\frac{1}{2\lambda-1},1\right\}  \right)\\
\label{eq:redu_I2rad}
I_{2}^{(rad)}(\lambda) &=& \frac{2^{2\lambda-3}Z^{3(\lambda-1)}}{n^{5\lambda-3}}\lambda^{-3}\, \Phi_{0}\left(2,0,2\lambda,\{n-1\},\{1\};\left\{\frac{1}{\lambda},1\right\}\right),
\end{eqnarray}
with 
\begin{eqnarray}
\mathcal{G}(n,0,\lambda)& =(2\lambda-1)^{-2} &\Bigg[ \Phi_{0}\left(2,0,2(2\lambda-1),\{n-1,\ldots,n-1\},\{1,\ldots,1\};\left\{\frac{1}{2\lambda-1},1\right\}\right)\nonumber \\
& & \hspace{-2.4cm}+4\,\Phi_{0}\Bigg(2,0,2(2\lambda-1),\{n-1,\ldots,n-1,n-2,n-2\},\{1,\ldots,1,2,2\};\left\{\frac{1}{2\lambda-1},1\right\}\Bigg)\nonumber\\
& & \hspace{-2.4cm}+4\,\Phi_{0}\Bigg(2,0,2(2\lambda-1),\{n-1,\ldots,n-1,n-2\},\{1,\ldots,1,2\};\left\{\frac{1}{2\lambda-1},1\right\}\Bigg) \Bigg].\nonumber\\
\end{eqnarray}
Thus, finally, the one-parameter ($\lambda$) Fisher-Rényi complexity measure $C^{(\lambda)}_{FR}[\rho_{ns}]$ for the $(ns)$-like hydrogenic states is given by
\begin{equation}
\label{eq:comp3}
C^{(\lambda)}_{FR}[\rho_{ns}] = D_{\lambda}^{-1}\frac{2^{3+\frac{2}{3 (\lambda -1)}}\pi^{\frac{2}{3}}}{ n^{-\frac{2}{3} \left(\frac{2}{\lambda -1}+5\right)} }\lambda ^{\frac{2}{\lambda -1}+6}(2 \lambda -1)^{-1}\mathcal{F}(n,0,\lambda), 
\end{equation}
where 
\begin{equation}
\label{eq:cons1}
\mathcal{F}(n,0,\lambda) = \Phi_{0}\left(2,0,2\lambda,\{n-1\},\{1\};\left\{\frac{1}{\lambda},1\right\}\right)^{2 \left(\frac{1}{3 (1-\lambda )}-1\right)}\mathcal{G}(n,0,\lambda).
\end{equation}
In particular, for the ground state (i.e., when $n=1, l=m=0$) we have shown in Appendix \ref{ground:app} that 

\begin{equation*}
\mathcal{F}(1,0,\lambda)  = 2^{2\left(\frac{1}{3(1-\lambda)}-1\right)}2(2\lambda-1)^{-2},
\end{equation*}
which allows us to find the following value 
\begin{equation}
\label{ground}
C^{(\lambda)}_{FR}[\rho_{1s}] = D_{\lambda}^{-1}4\pi^{\frac{2}{3}}\lambda^{\frac{2}{\lambda-1}+6}(2\lambda-1)^{-3}. 
\end{equation}
for the one-parameter Fisher-R\'enyi complexity measure of the hydrogenic ground state, keeping in mind the value (\ref{eq:normfactor}) for the normalization factor $D_{\lambda}$. We have done this calculation in detail to check our methodology; we are aware that in this concrete example it would have been simpler to start directly from the explicit expression of the wave function of the orbital $1s$. Operating in a similar way we can obtain the complexity values for the rest of $ns$-orbitals.
\vspace{3mm}

\item Generalized Fisher-Rényi complexity of hydrogenic circular states.

For these particular states the degree and parameter, $n-l-1$ and $2l+1$, of the orthonormal Laguerre polynomials, become $0$ and $2n-1$ respectively, so that the corresponding polynomials simplify as $\widehat{L}_{0}^{(2n-1)} (\tilde{r}) = \frac{1}{\sqrt{\Gamma(2n)}}$
and then the involved radial integrals follow as
\begin{eqnarray}
 \label{eq:I1a1}
 I_{1a}^{(rad)} &=& \int_{0}^{\infty} \left[\rho_{n,l}(\tilde{r})\right]^{2\lambda-3}\left[\frac{d}{dr}\rho_{n,l}(\tilde{r})\right]^{2}r^{2}\,dr \nonumber\\
&=& \frac{2^{4\lambda-3}Z^{6\lambda-4}}{n^{8\lambda-5}}\int_{0}^{\infty} \left\{[\widehat{L}_{n-l-1}^{(\alpha)}(\tilde{r})]^{2}\omega_{\alpha}(\tilde{r}) \right\}^{2\lambda-3}\left\{\frac{d}{d\tilde{r}}\left([\widehat{L}_{n-l-1}^{(\alpha)}(\tilde{r})]^{2}\frac{\omega_{\alpha}(\tilde{r})}{\tilde{r}} \right) \right\}^{2} \tilde{r}^{5-2\lambda}\, d\tilde{r}\nonumber\\
&=& \frac{2^{4\lambda-3}Z^{6\lambda-4}}{n^{8\lambda-5}}\left[\widehat{L}_{0}^{(2n-1)}\right]^{4\lambda-2}\int_{0}^{\infty} \omega_{2n-1}(\tilde{r})^{2\lambda-3}(\omega_{2n-1}^{'}(\tilde{r})\tilde{r}-\omega_{2n-1}(\tilde{r}))^{2} \,\tilde{r}^{1-2\lambda} \, d\tilde{r}\nonumber\\
&=& \frac{2^{2(2 \lambda -1)} Z^{2(3\lambda-2)}}{n^{8\lambda-5}} (2 \lambda  (n-1)-n+2) (2 \lambda -1)^{4 \lambda(1 - n)+2 n-5} \frac{\Gamma [3-2 n+4\lambda (n-1)]}{[\Gamma (2 n)]^{2 \lambda -1}},\nonumber \\
\\
 I_{1b}^{(rad)} &=& \int_{0}^{\infty} \left[\rho_{n,l}(\tilde{r})\right]^{2\lambda-1}\,dr  \nonumber\\ 
 &=& \frac{2^{4\lambda-3}Z^{2(3\lambda-2)}}{n^{8\lambda-5}}(2 \lambda -1)^{4 \lambda(1-n) +2 n-3}\,\frac{\Gamma[3-2n+4\lambda(n-1)]}{[\Gamma (2 n)]^{2\lambda-1}}.\\
 \label{eq:I2rad}
I_{2}^{(rad)} (\lambda)&=& \int_{0}^{\infty} \left[\rho_{n,l}(\tilde{r})\right]^{\lambda}r^{2}\,dr \nonumber \\
					&= & \frac{2^{2\lambda-3}Z^{3(\lambda-1)}}{n^{4\lambda-3}}\int_{0}^{\infty} \left\{[\widehat{L}_{n-l-1}^{(\alpha)}(\tilde{r})]^{2}\omega_{\alpha}(\tilde{r}) \right\}^{\lambda} \tilde{r}^{2-\lambda}\, d\tilde{r}\nonumber \\
					&=&\frac{2^{2\lambda-3}Z^{3(\lambda-1)}}{n^{4\lambda-3}}[\widehat{L}_{0}^{(2n-1)}]^{2\lambda}\int_{0}^{\infty}\omega_{2n-1}(\tilde{r})^{\lambda} \tilde{r}^{2-\lambda}\, d\tilde{r}  \nonumber \\
					&=& \frac{2^{2\lambda-3}Z^{3(\lambda-1)}}{n^{4\lambda-3}}[\widehat{L}_{0}^{(2n-1)}]^{2\lambda}\int_{0}^{\infty}e^{-\lambda\tilde{r}} \tilde{r}^{2(1+\lambda(n-1))}\, d\tilde{r}\nonumber \\
					&=& \frac{2^{2 \lambda -3} Z^{3(\lambda-1)}}{n^{4\lambda-3}} \lambda ^{-2 \lambda  (n-1)-3} \frac{\Gamma [2 (n-1) \lambda +3]}{ [\Gamma (2 n)]^{\lambda }},
\end{eqnarray}
On the other hand, the angular part of the wavefunction for the circular states reduces as
\begin{equation}
\label{eq:simplyspha}
\Theta_{n-1,n-1}(\theta,\phi) = |Y_{n-1,n-1}(\theta,\phi) |^{2} = \frac{\Gamma(n+1/2)}{2\pi^{3/2}\Gamma(n)}(\sin\theta)^{2(n-1)},
\end{equation}
which allows us to readily compute the angular integrals $I_{1a}^{(ang)}$, $I_{1b}^{(ang)}$ and $I_{2}^{(ang)}$ as 
\begin{eqnarray}
\label{eq:I1aang}
I_{1a}^{(ang)} &=& 2\pi\left[\frac{\Gamma(n+1/2)}{2\pi^{3/2}\Gamma(n)}\right]^{2\lambda-1}\int_{0}^{\pi} (\sin\theta)^{2(n-1)(2\lambda-1)}\sin\theta\, d\theta\nonumber \\
&=& 2^{2(1-\lambda)}\pi^{3(1-\lambda)}\left[\frac{\Gamma(n+1/2)}{\Gamma(n)}\right]^{2\lambda-1}\frac{\Gamma(2-n+2\lambda(n-1))}{\Gamma(5/2-n+2\lambda(n-1))}, \\
\label{eq:I2aang}
I_{1b}^{(ang)} &=& 2\pi\left[\frac{\Gamma(n+1/2)}{2\pi^{3/2}\Gamma(n)}\right]^{2\lambda-1}\int_{0}^{\pi} (\sin\theta)^{2(n-1)(2\lambda-3)}\left[\frac{d}{d\theta}(\sin\theta)^{2(n-1)}\right]^{2}\sin\theta\, d\theta\nonumber \\
&=& 2^{3-2 \lambda } \pi ^{3(1- \lambda) } (n-1)^2\left[\frac{\Gamma \left(n+\frac{1}{2}\right)}{\Gamma (n)}\right]^{2 \lambda -1}\frac{\Gamma[(2 \lambda -1) (n-1)]}{\Gamma \left[2 \lambda( n-1)-n+\frac{5}{2}\right]},  \\
\label{eq:I2ang}
I_{2}^{(ang)} &=& 2\pi \left[\frac{\Gamma(n+1/2)}{2\pi^{3/2}\Gamma(n)}\right]^{\lambda}\int_{0}^{\pi} (\sin\theta)^{2(n-1)\lambda}\sin\theta\, d\theta\nonumber \\
&=& 2^{1-\lambda}\pi^{\frac{3}{2}(1-\lambda)}\left[\frac{\Gamma(n+1/2)}{\Gamma(n)}\right]^{\lambda}\frac{\Gamma[1+\lambda(n-1)]}{\Gamma[\frac{3}{2}+\lambda(n-1)]}.
\end{eqnarray}
Gathering the last six numbered expressions together with Eqs. (\ref{eq:I1a}) and (\ref{eq:1B}), one finally obtains according to (\ref{eq:exp1}) the following value 
\begin{eqnarray}
\label{eq:circgencomp}
C^{(\lambda)}_{FR}[\rho_{cs}] &=& D_{\lambda}^{-1} \frac{2^{\frac{19}{3}-4 \lambda +\frac{2}{3 (\lambda -1)}+n(4 \lambda -2)}\pi ^{\frac{1}{2}}}{n^{\frac{2}{3 (1-\lambda)}-\frac{5}{3}} } \lambda ^{\frac{2 (3 \lambda -2) (2 \lambda  (n-1)+3)}{3 (\lambda -1)}} (2 \lambda -1)^{4 \lambda(1 -  n)+2 n-5}\nonumber\\
& &\hspace{-2.2cm}\times \frac{[\Gamma (n)\Gamma (2 n)]^{\frac{2}{3 (\lambda -1)}+\frac{5}{3}}\Gamma[2-n+2\lambda(n-1)]^2}{\Gamma \left(n+\frac{1}{2}\right)^{\frac{3-5 \lambda }{3 (1-\lambda )}} }\left[\frac{\Gamma \left(\frac{3}{2}+\lambda(n-1)\right)}{\Gamma (1+\lambda(n-1) )\Gamma(3+2\lambda(n-1))}\right]^{2\left(\frac{1}{3( \lambda -1)}+1\right)}.
\end{eqnarray}
\end{enumerate}
for the one-parameter Fisher-Rényi complexity measure of the hydrogenic circular states. This expression gives for the ground state (which is also a particular circular state with $l=n-1=0$) the same previously obtained value (\ref{ground}), what is a further checking of our results.

\section{Conclusions}

In this article we first explored a notion of complexity quantifier for the finite quantum many-particle systems, the one-parameter Fisher-Rényi complexity, and examined its main analytical properties. This notion extends all the previously known measures of complexity which are sensitive to the quantum fluctuations of the physical wavefunctions of the systems (Crámer-Rao, Fisher-Shannon, Fisher-Rényi-type) in the following sense: it does not depend on any specific point of the system's region (opposite to the Crámer-Rao measure) and it quantifies the combined balance of various aspects of the fluctuations of the single-particle density beyond the gradient content (opposite to the Fisher-Shannon complexity and the Fisher-Rényi product, which only take into account a single aspect given by the density gradient content) and different facets of the spreading of this density function.\\
Then, we illustrated the applicability of this novel measure of complexity in the main prototype of electronic systems, the hydrogenic atom. We have obtained an analytically, algorithmic way to calculate its values for all quantum hydrogenic states, and we have given the explicit values for all the $ns$ states and the circular states, which are specially relevant \textit{per se} and because they can be used as reference values for the complexity of Coulombian systems as reflected by the rich three-dimensional geometries of the electron density corresponding to their quantum states.

\acknowledgments{
 This work was partially supported by the Projects
P11-FQM-7276 and FQM-207 of the Junta de Andalucia, and by the MINECO-FEDER (European regional development fund) grants FIS2014- 54497P and FIS2014-59311P.
\L .R. acknowledges financial support by the grant number 2014/13/D/ST2/01886
of the National Science Center, Poland. Research in Cologne
is supported by the Excellence Initiative of the German Federal and State Governments (Grant
ZUK 81) and the DFG (GRO 4334/2-1). \L .R.  also acknowledges the support by the Foundation for Polish
Science (FNP) and hospitality of Freiburg Center for
Data Analysis and Modeling. I. V. Toranzo acknowledges the support of the Spanish Ministerio de Educación under the program FPU 2014.}

\appendix

\section{Calculation of the Fisher and Rényi-like hydrogenic integrals}
\label{integrals:app}
Let us here show the methodology to solve the integrals
 \begin{eqnarray}
 \label{eq:1A}
 I_{1} &=& \int |\left[\rho_{n,l,m}(\vec{r})\right]^{\lambda-2}\,\,\nabla\rho_{n,l,m}(\vec{r})|^{2}\,\rho_{n,l,m}(\vec{r})\, d\vec{r}=\int \left[\rho_{n,l,m}(\vec{r})\right]^{2\lambda-3}\,\,|\nabla\rho_{n,l,m}(\vec{r})|^{2}\,d\vec{r}, \\
 I_{2} &=&  \int \left[\rho_{n,l,m}(\vec{r})\right]^{\lambda}\, d\vec{r} =  I_{2}^{(rad)}\times I_{2}^{(ang)},
  \label{eq:1B}
 \end{eqnarray}
 with 
  \begin{eqnarray}
   \label{eq:reduc_I2}
  I_{2}^{(rad)}(\lambda) &=& \int_{0}^{\infty} \left[\rho_{n,l}(\tilde{r})\right]^{\lambda}r^{2}\,dr,
 \end{eqnarray} 
 and
  \begin{eqnarray}
   \label{eq:reduc_I2ang}
    I_{2}^{(ang)}(\lambda) &=& \int_{\Omega}\left[\Theta_{l,m}(\theta,\phi)\right]^{\lambda}\,d\Omega
 \end{eqnarray} 
 encountered in Section 3. Since the gradient operator is $\nabla = \left(\frac{\partial}{\partial r},\frac{1}{r}\frac{\partial}{\partial \theta}, \frac{1}{r\sin\theta}\frac{\partial}{\partial \phi} \right)$ and the probability density does not depend on the azimuthal angle, $\phi$, the integral $I_{1}$ can be written as
\begin{eqnarray}
\label{eq:I1a}
I_{1} &=& \int \left[\rho_{n,l,m}(\vec{r})\right]^{2\lambda-3}\,\left[\frac{\partial}{\partial r}\rho_{n,l,m}(\vec{r})\right]^{2}\, d^3\vec{r} + \int \left[\rho_{n,l,m}(\vec{r})\right]^{2\lambda-3}\,\left[\frac{1}{r}\frac{\partial}{\partial \theta}\rho_{n,l,m}(\vec{r})\right]^{2}\, d^3\vec{r} \nonumber \\
&\equiv & I_{1a}^{(rad)}\times I_{1a}^{(ang)}+I_{1b}^{(rad)}\times I_{1b}^{(ang)},
\end{eqnarray}
where one has used that $\frac{d}{dr} = \frac{2Z}{n}\frac{d}{d\tilde{r}}$, and
 \begin{eqnarray}
 \label{eq:reduc_I1a}
 I_{1a}^{(rad)} &=& \int_{0}^{\infty} \left[\rho_{n,l}(\tilde{r})\right]^{2\lambda-3}\left[\frac{d}{dr}\rho_{n,l}(\tilde{r})\right]^{2}r^{2}\,dr\\
 \label{eq:reduc_I1b}
  I_{1b}^{(rad)} &=& \int_{0}^{\infty} \left[\rho_{n,l}(\tilde{r})\right]^{2\lambda-1}\,dr ,
 \end{eqnarray} 
and 
\begin{eqnarray}
 \label{eq:reduc_I1ang}
 I_{1a}^{(ang)} &=& \int_{\Omega}\left[\Theta_{l,m}(\theta,\phi)\right]^{2\lambda-1}\,d\Omega \,\, = I_{2}^{(ang)}(2\lambda-1)  \\
 \label{eq:reduc_Ibang}
  I_{1b}^{(ang)} &=& \int_{\Omega}\left[\Theta_{l,m}(\theta,\phi)\right]^{2\lambda-3}\left[\frac{d}{d\theta}\Theta_{l,m}(\theta,\phi)\right]^{2}\,d\Omega,
 \end{eqnarray} 

Then, the complexity measure (\ref{eq:exp1}) can be rewritten as
  \begin{eqnarray}
  \label{eq:comp2}
  C^{(\lambda)}_{FR}[\rho_{n,l,m}] &=& D_\lambda^{-1} \left[I_{1a}^{(rad)}\times 	\left(I_{2}^{(rad)}\right)^{2 \left(\frac{1}{3 (1-\lambda )}-1\right)}\right]\left[I_{1a}^{(ang)}\times \left(I_{2}^{(ang)}\right)^{2 \left(\frac{1}{3 (1-\lambda )}-1\right)}\right]\nonumber \\
  & &+\left[I_{1b}^{(rad)}\times \left(I_{2}^{(rad)}\right)^{2 \left(\frac{1}{3 (1-\lambda )}-1\right)}\right]\left[I_{1b}^{(ang)}\times \left(I_{2}^{(ang)}\right)^{2 \left(\frac{1}{3 (1-\lambda )}-1\right)}\right]
  \end{eqnarray}
It remains to calculate the radial integrals $I_{1a}^{(rad)}$, $I_{1b}^{(rad)}$ and $I_{2}^{(rad)}$ and the angular integrals $I_{1a}^{(ang)}$, $I_{1b}^{(ang)}$ and $I_{2}^{(ang)}$. Let us start with the analytical determination of the radial integrals $I_{1}^{(rad)}$ and $I_{2}^{(rad)}$. To do that we use the differential relation of the Laguerre polynomials \cite{nist}
\begin{equation}
 \label{eq:tool1}
 \frac{d}{dx}L^{(\alpha)}_{n}(x) = -L_{n-1}^{(\alpha+1)}(x),
 \end{equation}
 and the linearization-like formula of Srivastava-Niukkanen \cite{srivastava,pablo} for the product of several Laguerre polynomials given by
 \begin{equation}
 \label{eq:tool2}
 x^{\mu}L_{m_{1}}^{(\alpha_{1})}(t_{1}x)\cdots L_{m_{r}}^{(\alpha_{r})}(t_{r}x) = \sum_{k=0}^{\infty}\Phi_{k}(\mu,\beta,r,\{m_{i}\},\{\alpha_{i}\};\{t_{i},1\})L_{k}^{(\beta)}(x)
 \end{equation}
 where the $\Phi_k$-linearization coeffients are 
 \begin{eqnarray}
 \label{eq:lauri}
 \Phi_{k}(\mu,\beta,r,\{m_{i}\},\{\alpha_{i}\};\{t_{i},1\}) &=& (\beta+1)_{\mu}\binom{m_{1}+\alpha_{1}}{m_{1}}\cdots \binom{m_{r}+\alpha_{r}}{m_{r}}\times\\
& & F_{A}^{r+1}(\beta+\mu+1, -m_{1},\ldots, -m_{r},-k;\alpha_{1}+1,\ldots,\alpha_{r}+1,\beta+1;t_{1},\ldots,t_{r},1)\nonumber 
 \end{eqnarray}
with the Pochhammer symbol \cite{nist} $(a)_{\mu}$, the binomial number $\binom{a}{b}$, and the Lauricella hypergeometric function of $(r+1)$ variables $F_{A}^{r+1}$ \cite{srivastava,pablo}.\\
Then, we obtain the following analytical expressions for the radial integrals in terms of the parameters $\{Z, \lambda, n, l\}$ of the system:
\begin{eqnarray}
\label{eq:Ia}
I_{1a}^{(rad)} &=& \frac{2^{4\lambda-3}Z^{6\lambda-4}}{n^{8\lambda-5}}\left[\frac{\Gamma(n-l)}{\Gamma(n+l+1)}\right]^{2\lambda-1}(2\lambda-1)^{-2l(2\lambda-1)-1}\mathcal{G}(n, l, \lambda),\\[1em] 
\label{eq:Ib}
I_{1b}^{(rad)} &=& \frac{2^{4\lambda-3}Z^{6\lambda-4}}{n^{8\lambda-5}}\left[\frac{\Gamma(n-l)}{\Gamma(n+l+1)}\right]^{2\lambda-1}(2\lambda-1)^{-2l(2\lambda-1)-1}\\
& & \times \Phi_{0}\left(2l(2\lambda-1),0,2(2\lambda-1),\{n-l-1\}, \{2l+1\};\left\{\frac{1}{2\lambda-1},1\right\}  \right),\nonumber\\[1em]
\label{eq:I2}
I_{2}^{(rad)}(\lambda) &=&  \frac{2^{2\lambda-3}Z^{3(\lambda-1)}}{n^{4\lambda-3}}\left[\frac{\Gamma(n-l)}{\Gamma(n+l+1)}\right]^{\lambda}\lambda^{-2l\lambda-3}\nonumber\\ & & \times \Phi_{0}\left(2(l\lambda+1),0,2\lambda,\{n-l-1\},\{2l+1\};\left\{\frac{1}{\lambda},1\right\}\right), \nonumber  \\
\end{eqnarray}
where $\mathcal{G}(n, l, \lambda)$ is
\begin{eqnarray}
\label{eq:funcG}
\mathcal{G}(n, l, \lambda) & =& \Bigg[4l^{2} \Phi_{0}\left(2l(2\lambda-1),0,2(2\lambda-1),\{n-l-1,\ldots,n-l-1\},\{2l+1,\ldots,2l+1\};\left\{\frac{1}{2\lambda-1},1\right\}\right) \nonumber\\
& & + (2\lambda-1)^{-2}\nonumber\\
& &\times\Phi_{0}\left(2l(2\lambda-1)+2,0,2(2\lambda-1),\{n-l-1,\ldots,n-l-1\},\{2l+1,\ldots,2l+1\};\left\{\frac{1}{2\lambda-1},1\right\}\right)\nonumber \\
& &-4l(2\lambda-1)^{-1}\nonumber\\ & & \times \Phi_{0}\left(2l(2\lambda-1)+1,0,2(2\lambda-1),\{n-l-1,\ldots,n-l-1\},\{2l+1,\ldots,2l+1\};\left\{\frac{1}{2\lambda-1},1\right\}\right) \nonumber \\
& &+\frac{4}{(2\lambda-1)^{2}}\times\nonumber\\
& & \Phi_{0}\Bigg(2l(2\lambda-1)+2,0,2(2\lambda-1),\{n-l-1,\ldots,n-l-1,n-l-2,n-l-2\},\nonumber\\
& & \{2l+1,\ldots,2l+1, 2l+2,2l+2\};\left\{\frac{1}{2\lambda-1},1\right\}\Bigg)\nonumber\\
& & -\frac{8l}{(2\lambda-1)}\times \nonumber \\ & &\Phi_{0}\Bigg(2l(2\lambda-1)+1,0,2(2\lambda-1),\{n-l-1,\ldots,n-l-1, n-l-2\},\nonumber\\
& &\{2l+1,\ldots,2l+1,2l+2\};\left\{\frac{1}{2\lambda-1},1\right\}\Bigg) \nonumber \\
& & +\frac{4}{(2\lambda-1)^{2}}\times \nonumber \\  && \Phi_{0}\Bigg(2l(2\lambda-1)+2,0,2(2\lambda-1),\{n-l-1,\ldots,n-l-1,n-l-2\},\nonumber\\
& & \{2l+1,\ldots,2l+1,2l+2\};\left\{\frac{1}{2\lambda-1},1\right\}\Bigg) \Bigg],
\end{eqnarray}
where one should keep in mind that the $\Phi_{0}$ functions are given as in (\ref{eq:lauri}).\\
Similarly we can obtain the angular integrals by means of linerization-like formulas of the Gegenbauer polynomials or the associated Legendre polynomials of the first kind.

\section{Calculation of $\mathcal{F}(1,0,\lambda)$}
\label{ground:app}

Here we will determine the value of 
\begin{equation*}
\mathcal{F}(1,0,\lambda) = \Phi_{0}\left( 2,0,2\lambda,\{0\},\{1\};\left\{\frac{1}{\lambda},1 \right\} \right)^{2\left(\frac{1}{3(1-\lambda)}-1 \right)}\mathcal{G}(1,0,\lambda)
\end{equation*}
where
\begin{eqnarray*}
\Phi_{0}\left( 2,0,2\lambda,\{0\},\{1\};\left\{\frac{1}{\lambda},1 \right\} \right) &=& (1)_{2}\binom{1}{0}^{2\lambda} F_{A}^{2\lambda+1}\left(3, 0, \ldots, 0, 0; 2, \ldots, 2,1; \frac{1}{\lambda},\ldots,\frac{1}{\lambda},1 \right)\\
&  & \hspace{-5cm}=\sum_{j_{1},\ldots,j_{2\lambda+1}=0}^{\infty} \frac{(3)_{j_{1}+\ldots+j_{2\lambda+1}}(0)_{j_{1}}\ldots(0)_{j_{2\lambda+1}}}{(2)_{j_{1}}\ldots(2)_{j_{2\lambda+1}}}\left(\frac{1}{\lambda}\right)^{j_{1}+\ldots+j_{2\lambda+1}}\frac{1}{j_{1}!\ldots j_{2\lambda+1}!} =2 
\end{eqnarray*}
and
\begin{eqnarray*}
\mathcal{G}(1,0,\lambda) &=& (2\lambda-1)^{-2}\Bigg[ \Phi_{0}\left( 2,0,2(2\lambda-1),\{0\},\{1\};\left\{\frac{1}{2\lambda-1},1 \right\} \right) \\
& & +4 \Phi_{0}\left( 2,0,2(2\lambda-1),\{0,\ldots,0,-1,-1\},\{1,\ldots,1,2,2\};\left\{\frac{1}{2\lambda-1},1 \right\} \right)\\
& & + 4\Phi_{0}\left( 2,0,2(2\lambda-1),\{0,\ldots,0,-1\},\{1,\ldots,1,2\};\left\{\frac{1}{2\lambda-1},1 \right\} \right)\Bigg]\\
&=& (2\lambda-1)^{-2} [2 + 4\cdot 0 + 4\cdot 0 ] = 2 (2\lambda-1)^{-2}
\end{eqnarray*}
since 
\begin{eqnarray*}
\Phi_{0}\left( 2,0,2(2\lambda-1),\{0,\ldots,0,-1\},\{1,\ldots,1,2\};\left\{\frac{1}{2\lambda-1},1 \right\} \right) & &
\\
= (1)_{0}\binom{1}{0}^{2(2\lambda-1)-1}\binom{1}{-1}F_{A}^{2(2\lambda-1)+1}(\ldots) =0.   & &
\end{eqnarray*}
Then, we obtain that
\begin{equation*}
\mathcal{F}(1,0,\lambda)  = 2^{2\left(\frac{1}{3(\lambda-1)}-1\right)}2(2\lambda-1)^{-2}.
\end{equation*}






\end{document}